\documentclass[reviewcopy]{elsarticle}
\usepackage{booktabs}
\usepackage[reviewcopy]{adndt}
\usepackage{longtable}
\usepackage{graphicx}
\usepackage{epsfig}
\usepackage{epstopdf}
\usepackage{multicol}
\usepackage{multirow}
\usepackage[toc]{appendix}
\usepackage[dvipsnames]{xcolor}

\usepackage{amsmath}

\usepackage{amssymb}
\usepackage{comment}
\usepackage{multirow}
\usepackage{tabularray}
\usepackage{xcolor}

\biboptions{square,sort&compress}
\bibpunct[]{[}{]}{,}{n}{}{;}
\begin{document}

\begin{frontmatter}

\journal{Atomic Data and Nuclear Data Tables}
\title{Bulk and neutron-proton asymmetry coefficients of the semi-empirical mass formula tuned to ground state mass excess of AME2020 and/or FRDM(2012)}
\author{Dalip Singh Verma\corref{cor1}}
  \ead{dsverma@cuhimachal.ac.in}
  \author{Vivek}
  \author{Kushmakshi}
  \cortext[cor1]{Corresponding author.}
  \address{Department of Physics and Astronomical Science\\ Central University of Himachal Pradesh, Dharamshala, District Kangra (H.P.)-176215}
\date{16.12.2002} 
\begin{abstract}
Davidson $\emph{et al.}$ has extended Seeger's mass formula to non-zero excitation energies by introducing temperature-dependent coefficients in the liquid drop energy part of the semi-empirical mass formula, without considering the shell effects. The semi-empirical mass formula of Davidson $\emph{et al.}$ is applicable for the compound nucleus temperatures less than or equal to 4 MeV. The mass excess calculated using this mass formula including shell effects/corrections does not reproduce the ground state mass excesses of the new atomic mass evaluation data AME2020 and/or FRDM(2012) with its coefficients at zero temperature. So, the coefficients of the semi-empirical mass formula are needed to be tuned to reproduce the ground state mass excess of the nuclei in the recent atomic mass evaluation data AME2020 and/or FRDM(2012). The bulk and neutron-proton asymmetry coefficients of the semi-empirical mass formula of Davidson $\emph{et al.}$ with shell effects have been tuned to reproduce the mass excess data for all the nuclei of AME2020 (Z=1-118 and A=1-295)  and the nuclei of FRDM(2012) (Z=8-136 and A=16-339, except 3456 nuclei which are also available in the AME2020 data) at zero temperature, i.e., the coefficients are tuned for 9420 nuclei known at present.  The tuned bulk and neutron-proton asymmetry coefficients reproduce the mass excess of the new atomic mass evaluation data AME2020 and/or FRDM(2012) within a difference of less than 1 MeV and can be used for the applications/investigations in the areas of physics where high energies are experienced or nuclei involved are in excited states, e.g., fusion-evaporation and fusion-fission processes in heavy-ion reactions.
\end{abstract}
\end{frontmatter}
\newpage
\tableofcontents
\listoffigures
\listoftables
\vskip5pc
\section{Introduction}
Recently, new mass excess data is available for 3558 nuclei in the new atomic mass evaluation AME2020 \cite{AME2020} with nuclei of atomic and mass numbers ranging from Z=1-118 and A=1-295, respectively. It contains the mass excess data for 122 new nuclei that appeared in the atomic mass evaluation data AME2020 in addition to the revised mass excess data for 3436 nuclei of AME2016 \cite{AME2017}  and hence superseded the mass excess data of AME2016. In addition to the newly measured ground state mass excess data of AME2020, there is a new theoretical ground state mass excess data available for 9318 nuclei in FRDM(2012) \cite{FRDM2012} with nuclei of atomic and mass numbers ranging from Z=8-136 and A=16-339, respectively. The  FRDM(2012) contains the mass excess data of 339 new nuclei and revised mass excess data for 8979 nuclei of its previous generation FRDM(1992) mass table in M$\Ddot{o}$ller \emph{et al.} (1995) \cite{M1995}. The available data can be used to study the area of physics where the nuclei involved are in their ground state, like cluster radioactivity, spontaneous fission, etc. In the areas of physics like heavy-ion reactions and  astrophysics etc., the nuclei happen to be in their excited states and hence there is a need for mass excesses at higher energies or excitation energies.  In this direction, Davidson \emph{et al.} \cite{Dav1994} has extended Seeger's mass formula \cite{Seeger1961} by introducing the temperature-dependent coefficients to the liquid drop energy, excluding the shell corrections. The shell corrections have been introduced by Myers and Swiatecki \cite{Myers1966} and extended for the finite nuclear excitation energy by Jensen \emph{et al.} \cite{jensen73}. The mass excess at temperature T is expressed as $\Delta M(A, Z, T)=M(Z, A, T)-A=ZM_H+NM_n-B(A, Z, T)-A$, where the binding energies $B(A, Z, T)$ are expressed as the sum of temperature-dependent terms (i) liquid drop energies $V_{LDM}(T)$ of Davidson \emph{et al.} \cite{Dav1994} and (ii) shell corrections $\delta U (T)$, within the use of Strutinsky renormalization procedure \cite{VM}. But, this formula neither reproduces the recent ground state mass excess data of AME2020 nor of  FRDM(2012), see Fig. (\ref{MEZ8}) below. So, there is a need to adjust some of the coefficients of the semi-empirical mass formula of Davidson \emph{et al.} of Eq. (\ref{Eq.3}) at T=0 to reproduce the ground state mass excess data of AME2020 and/or FRDM(2012). Gupta \emph{et. al.} \cite{DCM2003} found that the measured ground state mass excess data can be fitted within 1–1.5 MeV by changing the bulk $\alpha$(0) and the neutron-proton asymmetry $a_a$ coefficients of the semi-empirical mass formula of Davidson \emph{et al.} at T=0 MeV, where the neutron-proton asymmetry coefficient $a_a$ controls the curvature of the measured mass excess parabola and the bulk coefficient $\alpha(0)$ acts as a scaling factor. First, these authors reproduced the ground state mass excess data of Audi and Wapstra AME1995 \cite{Audi1995} by adjusting these coefficients up to $Z$ = 28 \cite{DCM2003} and used the mass excess data to study the decay of hot $^{56}$Ni$^*$ formed in the $^{32}$S + $^{24}$Mg reaction \cite{DCM2003,Gupta2005}. Later they extended the fitting of these coefficients up to $Z$=56 and used them to study the emission of intermediate mass fragments from hot $^{116}$Ba$^*$ formed in low-energy $^{58}$Ni + $^{58}$Ni reaction, see ref. \cite{Gupta2003}. The adjustment of these coefficients has been extended further by \cite{Birbikram2008} up to Z=97 to reproduce the experimental mass excess of AME2003 \cite{A2003} and/or theoretical mass excess data of FRDM(1992) mass table in M$\Ddot{o}$ller \emph{et al.} (1995), within a difference of less than 1.5 MeV between the calculated ground state mass excess and the mass excess of AME2003 and/or FRDM(1992), and used for the decay study of hot compound nucleus $^{246}$Bk$^*$ at different excitation energies. The adjustment of these coefficients is then extended to Z=118 by \cite{Birbikram2008a}. Due to the addition of mass excess data for 257 new nuclei in AME2016  and 339 new nuclei FRDM(2012) with respect to the AME2003 and FRDM(1992), respectively, the coefficients  are adjusted by \cite{Thesis} for these new nuclei and used for the investigation of the cold valley paths for the synthesis of isotopes of Ubh in optimum orientations \cite{dalip2021}. It may be noted that a lot of work has been done using the dynamical cluster-decay model developed for the study of the decay of hot compound nuclei where the mass excess data of nuclei in excited states is used. With the availability of new mass excess data in AME2020 and FRDM(2012) the tuning of the bulk and neutron-proton asymmetry coefficients is needed again to reproduce the latest ground state mass excess data.

In this work, the bulk and neutron-proton asymmetry coefficients have been adjusted to reproduce the mass excess data for 122 new nuclei that appeared in AME2020 and readjusted the coefficients for the rest of the nuclei available in AME2020 and/or FRDM(2012) within a difference of less than 1 MeV. In total, we have obtained the bulk and neutron-proton asymmetry coefficients for all known 9420 nuclei (as listed in Table. \ref{Tab 1}) up to Z=136, where 3558 nuclei are the nuclei available in AME2020 of Z=1-118 and 5862 nuclei of FRDM(2012) of Z=8-136, but other than the nuclei available in AME2020. Using the new bulk and neutron-proton asymmetry coefficients listed in Table \ref{Tab 1} in the mass excess formula of Eq. (\ref{Eq.3}) one can calculate the mass excess for the nuclei at higher energies. It may be noted that some of the experimental generations of atomic mass evaluation are AME2003 (3179 nuclei), AME2012 (=AME2003+174 new nuclei) \cite{AME2012}, AME2016 (=AME2012+83 new nuclei) and AME2020 (=AME2016+122 new nuclei) and that of the theoretical one are FRDM(1992) (5523 nuclei) and FRDM(2012) (=FRDM(1992)+339 new nuclei).

The paper is organized as follows: Section \ref{M} describes the  mass excess formulae for the ground and excited states of nuclei, and the calculation detail of shell corrections. Section \ref{R} contains the calculations and discussion of the results and Section \ref{L} is the list of the present tuned bulk and neutron-proton asymmetry coefficients and the calculated mass excess compared with the mass excesses of AME2020 and/or FRDM(2012) data.

\section{Mass excess for nuclei in ground and excited states}\label{M}
The standard form of the mass excess formula of P. A. Seeger \cite{Seeger1961} for an atom of Z proton and N neutron is
\begin{eqnarray}\label{Eq.1}
\Delta M(Z,A)=M(Z,A)-A&=& M_{n}N+M_{H}Z-A-\alpha A + \left(\beta	-\frac{\eta}{A^{1/3}}\right)\left(\frac{I^2+2|I|}{A}\right)+\gamma A^{2/3} \nonumber \\
&+&\frac{3}{5}\frac{e^2}{R_0}\frac{Z^2}{A^{1/3}}\left(1-\frac{0.7636}{Z^{2/3}}-\frac{2.29}{R_0^{2} A^{2/3}}\right)+a\delta_a
\end{eqnarray}
where the mass excesses for neutron and hydrogen atom is $M_n-1$(=8.3674 MeV/c$^2$)  and $M_H-1$(=7.5848 MeV/c$^2$), respectively, with $A$ (=$N+Z$) the mass number of a nucleus, $I (=N-Z)$ is the neutron-proton asymmetry and the free coefficients  $\alpha$, $\beta$, $\eta$, and $\gamma$ respectively are 16.11, 20.65, 48.00, and 20.21 MeV/c$^2$. The R$_0$ is 1.07 fm, $\delta_a$ is the pairing term and the coefficient of the pairing term $a$ is 1/2, 0, -1/2 for odd atoms, odd mass atoms, and even atoms, respectively. The mass excess formula, Eq. (\ref{Eq.1}), is extended by Davidson \emph{et. al.}  for hot nuclei by introducing the temperature-dependent coefficients  $\alpha(T)[=\alpha(0)+\frac{1}{15}T^2]$, $\beta(T)$, $\eta(T)$, $\gamma(T)$ and $\delta(T)$ \cite{Dav1994} and nuclear radius R$(T)$ $=R_0(1+0.01T)$ \cite{Brack1974} in liquid drop energy as
\begin{eqnarray}\label{Eq.2}
 V_{LDM}(Z,A,T)&=&\alpha(T)A+\biggr(\beta(T)-\frac{\eta(T)}{A^{1/3}}\biggl)\biggr(\frac{4t_{\zeta}^{2}+4|t_{\zeta}|}{A}\biggl)+\gamma(T)A^{2/3} \nonumber \\
 &+&\frac{Z^{2}}{R(T)A^{1/3}}\biggr(1-\frac{0.7636}{Z^{2/3}}-\frac{2.29}{[R(T)A^{1/3}]^{2}}\biggl)+\delta(T)\frac{f(Z,A)}{A^{3/4}}
\end{eqnarray}
where, $t_{\zeta}={a_a}(Z-N)$ with the neutron-proton asymmetry coefficient  $a_{a}=1/2$ and the function $f(Z,A)$ in pairing term is $f(Z,A)=(-1, 0, 1)$, respectively for even-even, odd-even and odd-odd nuclei. At T=0 the coefficients $\alpha$, $\beta$, $\eta$, and $\gamma$  are the same as given in Seeger's mass formula and the pairing term coefficient $\delta$ is 33.0 MeV \cite{DeBenedetti1964}, here $\alpha(0)$ = -16.11 MeV/c$^2$. For microscopic effects in liquid drop energy, the shell correction to the mass excess in the ground state (T=0) of a nucleus with the mass number $A$, neutron number $N$ and proton number $Z$  is of Myers and Swiatecki \cite{Myers1966}, given as
\begin{eqnarray}\label{Eq.4}
\delta U=C\biggr[\frac{F(N)+F(Z)}{(A/2)^{\frac{2}{3}}}-cA^{\frac{1}{3}}\biggl];\quad  F(X)=\frac{3}{5}\biggr(\frac{M^{{5}/{3}}_{i}-M^{{5}/{3}}_{i-1}}{M_{i}-M_{i-1}}\biggl)(X-M_{i-1})-\frac{3}{5}\biggr(X^{{5}/{3}}-M^{{5}/{3}}_{i-1}\biggl)
\end{eqnarray}
with $X = N ~\text{or} ~Z,~ M_{i-1} < X < M_{i}$ with $M_{i-1}$ and $M_{i}$ are the lower and upper limits of magic numbers for a set of nuclei of magic numbers 2, 8, 14 \cite{Fridmann2005} or 20, 28, 50, 82, 126 are both for proton and neutron while 184 and 258 \cite{shellclosure258} are for neutron only with $c$ = 0.26  and  $C$ = 5.8 MeV. The temperature-dependence in the shell corrections can be included using $\delta U (T)=\delta U\exp{(-T^2/T_0})$ with T$_0$=1.5 MeV \cite{jensen73}.  The temperature T of the compound nucleus of mass A can be obtained from the relation $E^*_{CN}=aT^2-T$, where $a$ is the level density parameter which is related to the mass of the compound nucleus of mass A as $a=A/9$ (for the heavy mass region) and $A/10$ (for the superheavy mass region), etc., and $E^*_{CN}=E_{cm}+Q_{in}$ is the compound nucleus excitation energy.
Using Eq. (\ref{Eq.2}) for temperature-dependent liquid drop energy V$_{LDM}(T)$ and shell correction $\delta U (T)$, the temperature-dependent mass excess formula becomes
\begin{eqnarray} \label{Eq.3}
  \Delta M(Z,A,T)&=&M_{n}N+M_{H}Z-A+\alpha(T) A+\biggr(\beta(T)-\frac{\eta(T)}{A^{1/3}}\biggl)\biggr(\frac{4t_{\zeta}^{2}+4|t_{\zeta}|}{A}\biggl)+\gamma(T) A^{2/3}\nonumber \\
  &+&\frac{Z^{2}}{R(T) A^{1/3}} \biggr(1-\frac{0.7636}{Z^{2/3}}-\frac{2.29}{[R(T) A^{1/3}]^{2}}\biggl)+\delta(T)\frac{f(Z,A)}{A^{3/4}}+\delta U(T)
\end{eqnarray}
The bulk $\alpha(0)$ and neutron-proton asymmetry coefficients a$_a$ of $t_{\zeta}[={a_a}(Z-N)]$ of Eq. (\ref{Eq.3}) are adjusted to reproduce the mass excess data of  AME2020 and/or theoretical FRDM(2012).
\section{Calculations and discussion of results}\label{R}
\begin{figure}[h]
\centering	
\includegraphics[width=0.755\textheight, keepaspectratio]{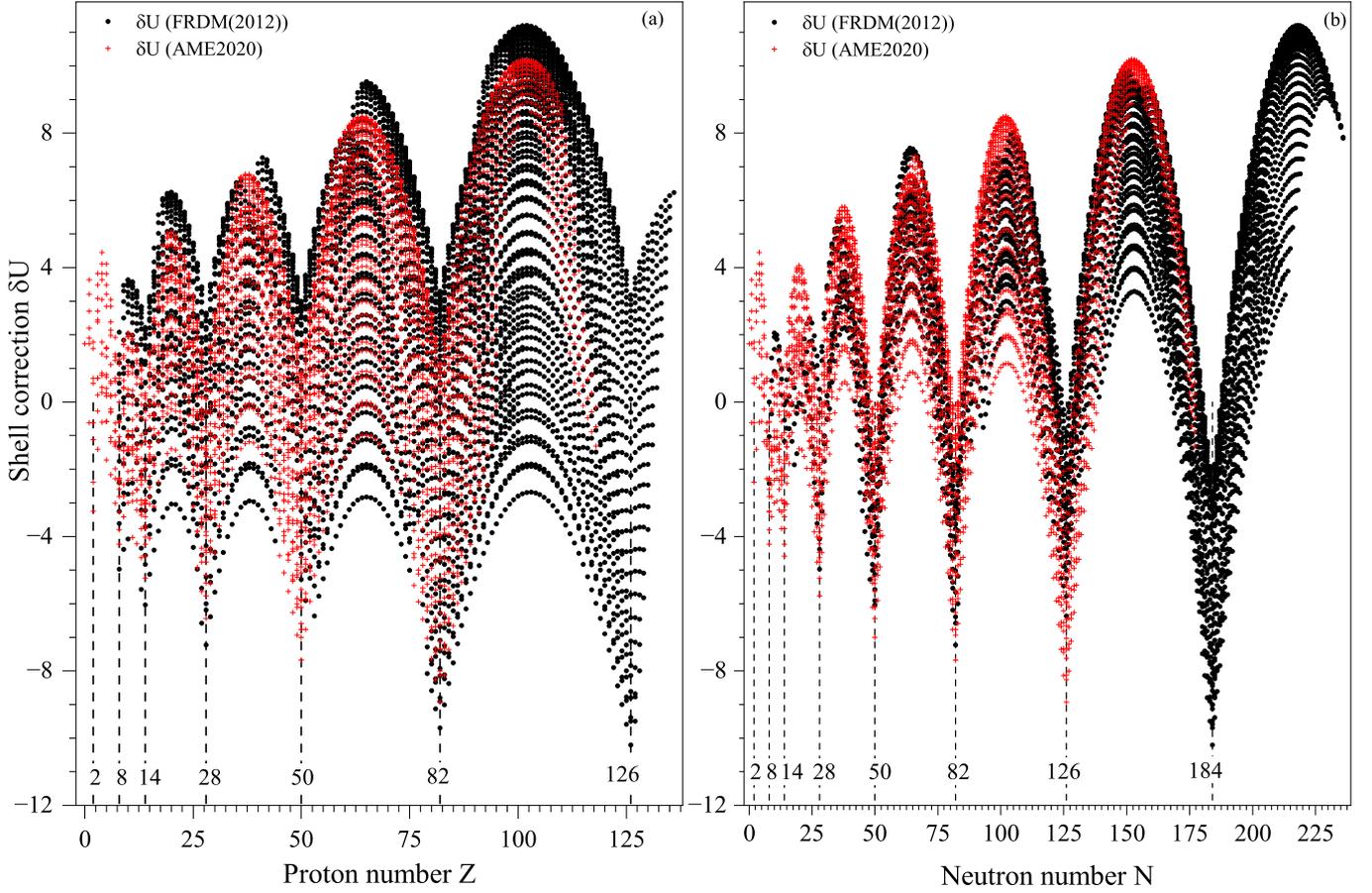}
\caption{The shell corrections calculated using Eq. (\ref{Eq.4}) for all the nuclei of AME2020 \cite{AME2020} and/or theoretical FRDM(2012) \cite{FRDM2012}, i.e., for all the nuclei listed in Table \ref{Tab 1}, as a function of (a) proton number $Z$ and (b) neutron number $N$ .}
\label{SC}
\end{figure}
Fig. (\ref{SC}) shows the calculated shell corrections for the 9420 nuclei of AME2020 and/or FRDM(2012) data as a function of (a) the proton number $Z$ and (b) the neutron number $N$, respectively. The $+$ symbol (red) is for the shell corrections of the nuclei of AME2020 data while the black solid circle $\bullet$ symbol is for the nuclei of FRDM(2012), which are not available in AME2020 data. The  vertical dotted lines show the position of the various magic numbers with respect to which the shell corrections have been calculated using Eq. (\ref{Eq.4}).

\begin{figure}[h]
\centering	
\includegraphics[width=0.755\textheight, keepaspectratio]{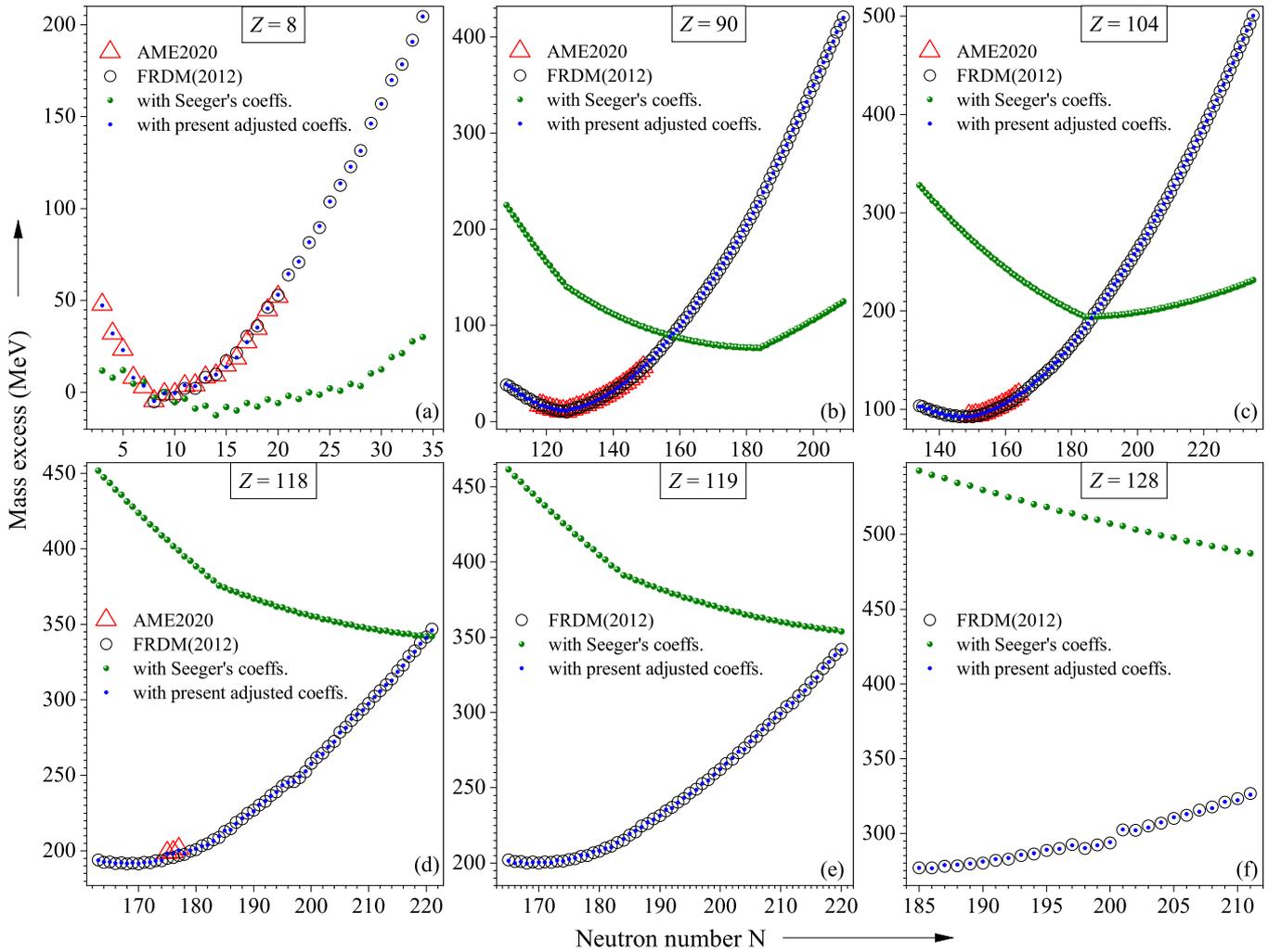}
\caption{The ground state mass excess for the isotopes of atomic number (a)  8,  (b)  90, (c) 104, (d) 118 (e) 119 and (f) 128 of (i) AME2020 \cite{AME2020} (ii) FRDM(2012) \cite{FRDM2012} (iii) and the calculated ground state mass excess of Eq. (\ref{Eq.3}) with (a) the coefficients of Seeger's mass formula (b) with present adjusted bulk and neutron-proton asymmetry coefficients of Eq. (\ref{Eq.3}), as a function of neutron number $N$.}
\label{MEZ8}
\end{figure}
 Fig. \ref{MEZ8} shows that the calculated mass excess using the mass excess formula of Eq. (\ref{Eq.3})  at T=0 MeV with Seeger's coefficients, shown by scattered green solid sphere $\bullet$ symbols, neither reproduce the ground state mass excess of AME2020 nor of the FRDM(2012), shown for the isotopes of Z = 8, 90, 104, 118, 119 and 128 (arbitrarily chosen Z of light, heavy and superheavy mass regions). As stated in the introduction, the measured ground state mass excess data can be fitted within 1–1.5 MeV by changing the bulk $\alpha$(0) and the neutron-proton asymmetry a$_a$ coefficients of the semi-empirical mass formula of Davidson \emph{et. al.} at T=0 MeV. This justifies the need of adjusting the bulk and neutron-proton asymmetry coefficients of the liquid drop energy of Davidson's \emph{et. al.} at T=0 MeV in the mass excess formula of Eq. ($\ref{Eq.3}$). The bulk and neutron-proton asymmetry coefficients of Eq. (\ref{Eq.3}) at T=0 are adjusted to reproduce the mass excess data of  AME2020 and/or FRDM(2012) within a deviation of less than 1 MeV. So, Eq. (\ref{Eq.3}) can be used to obtain the T-dependent mass excess for studying the decay/formation of hot compound nuclei, where according to the quantum mechanical fragmentation theory \cite{HJ,JM,AS} the fragmentation potential for a hot compound nucleus is expressed in terms of the temperature-dependent quantities like binding energies or mass excess, the proximity potential, the Coulomb potential and the centrifugal potentials of the fragments, see ref. \cite{Gupta2005} for detail. The present adjusted bulk and neutron-proton asymmetry coefficients reproduce the mass excess values of AME2020 and FRDM(2012) data within a difference of less than 1 MeV as shown in Fig. \ref{MEZ8} with symbol $\bullet$ (blue) for the isotopes of atomic number (a)  8,  (b)  90, (c) 104, (d) 118 (e) 119 and (f) 128. Here, the symbol $\Delta$ is for the mass excess of AME2020 and the symbol $\circ$ is for the FRDM(2012) mass excess. The mass excess calculated for the nuclei of AME2020 and/or FRDM(2012) is listed in Table \ref{Tab 1} below along with the adjusted bulk and neutron-proton asymmetry coefficients of the Eq. (\ref{Eq.3}) and comparison with the mass excess of AME2020 and/or FRDM(2012).

\section{The bulk and neutron-proton asymmetry coefficients and mass excesses}\label{L}
\fontsize{5.9}{5.9}\selectfont
\begin{longtable}{p{0.08cm}lp{0.81cm}lll|p{0.08cm}lp{0.81cm}lll|p{0.08cm}lp{0.81cm}lll}
\caption{ List of adjusted bulk $\alpha(0)$ and neutron-proton asymmetry $a_{a}$ coefficients and our calculated mass excess (Cal.) with the help of these adjusted coefficients compared with the experimental and theoretical mass excess of the nuclei of AME2020 \cite{AME2020} (Ex.) and/or FRDM(2012) \cite{FRDM2012} (Th., shown in \textbf{bold text})  \\ }
\label{Tab 1}\\
\hline
   &             &                  &                &    \multicolumn{2}{c|}{Mass excess (MeV)} &             &             &                  &                & \multicolumn{2}{c|}{Mass excess (MeV)} &             &             &                  &                & \multicolumn{2}{c}{Mass excess (MeV)} \\
   \cline{5-6}\cline{11-12}\cline{17-18}
\endfirsthead
\multicolumn{18}{c}%
{{\bfseries Table \thetable\ continued from previous page}} \\
\hline
            &             &                  &                & \multicolumn{2}{c|}{Mass excess (MeV)} &             &             &                  &                & \multicolumn{2}{c|}{Mass excess (MeV)} &             &             &                  &                & \multicolumn{2}{c}{Mass excess (MeV)} \\
             \cline{5-6}\cline{11-12}\cline{17-18}
\textbf{Z}  & \textbf{N}  & $\alpha(0)$      & a$_{a}$        & Cal.          & Ex./Th.            & \textbf{Z}  & \textbf{N}  & $\alpha(0)$      & a$_{a}$        & Cal.          & Ex./Th.            & \textbf{Z}  & \textbf{N}  & $\alpha(0)$      & a$_{a}$        & Cal.          & Ex./Th.            \\
\hline
\endhead
\textbf{Z}  & \textbf{N}  & $\alpha(0)$      & a$_{a}$        & Cal.          & Ex./Th.            & \textbf{Z}  & \textbf{N}  & $\alpha(0)$      & a$_{a}$        & Cal.          & Ex./Th.            & \textbf{Z}  & \textbf{N}  & $\alpha(0)$      & a$_{a}$        & Cal.          & Ex./Th.            \\
\hline
\textbf{1}  & 2           & -15.850     & 0.100   & 14.981  & 14.950           & \textbf{2}  & 1           & -15.600     & 0.070   & 14.863  & 14.931           & \textbf{3}  & 0           & -9.730      & 0.100   & 28.673  & 28.670           \\
            & 3           & -17.100     & 0.120   & 25.218  & 24.620           &             & 2           & -16.110     & 0.100   & 2.497   & 2.425            &             & 1           & -16.670     & 0.100   & 25.403  & 25.320           \\
            & 4           & -13.370     & 0.100   & 32.895  & 32.890           &             & 3           & -16.900     & 0.300   & 11.414  & 11.231           &             & 2           & -17.000     & 0.100   & 11.904  & 11.680           \\
            & 5           & -13.570     & 0.120   & 41.848  & 41.880           &             & 4           & -14.230     & 0.300   & 17.519  & 17.592           &             & 3           & -18.430     & 0.990   & 14.042  & 14.087           \\
            & 6           & -11.490     & 0.100   & 49.078  & 49.140           &             & 5           & -14.230     & 0.300   & 26.475  & 26.073           &             & 4           & -16.670     & 0.100   & 15.335  & 14.907           \\
            &             &             &         &         &                  &             & 6           & -13.120     & 0.100   & 31.588  & 31.610           &             & 5           & -16.670     & 0.100   & 20.277  & 20.946           \\
\hline
\end{longtable}
\fontsize{11}{11}\selectfont
\section*{Acknowledgments}
The authors are thankful to the Central University of Himachal Pradesh, Dharamshala, District Kangra (H.P.) for providing the necessary facility for completing this work.
\section*{Declarations}
The authors declare that they have no known competing financial interests or personal relationships that could have appeared to influence the work reported in this paper.

\clearpage

\end{document}